\documentclass[aps,amsmath,amssymb,showpacs,showkeys]{revtex4}
\usepackage[dvips]{graphicx,color}
\usepackage{times}
\usepackage{xcolor}
\usepackage{hyperref}
\hypersetup{colorlinks=true, citecolor=blue, filecolor=red, urlcolor=cyan}

\begin{document}
\title{Scalarons mimicking Dark Matter in the Hu-Sawicki model of $\mathbf{f(R)}$ gravity}

\author{Nashiba Parbin}
\email[Email: ]{nashibaparbin91@gmail.com}
\author{Umananda Dev Goswami}
\email[Email: ]{umananda2@gmail.com}
\affiliation{Department of Physics, Dibrugarh University, Dibrugarh 786004, 
Assam, India}

\begin{abstract}
In this paper, we conduct a study on the scalar field obtained from 
$\mathit{f(R)}$ gravity 
via Weyl transformation of the spacetime metric $g_{\mu\nu}$ from the Jordan 
frame to the Einstein frame. The scalar field is obtained as a result of the 
modification in the geometrical part of Einstein's field equation of 
General Relativity. For the Hu-Sawicki model of $\mathit{f(R)}$ gravity, we 
find the effective potential of the scalar field and calculate its mass. Our 
study shows that the scalar field (also named as scalaron) obtained from this 
model has the chameleonic property, i.e.\ the scalaron becomes light in the 
low-density region while it becomes heavy in the high-density region of matter.
Then it is found that the scalaron can be regarded as a dark matter (DM) 
candidate since the scalaron mass is found to be quite close to the mass of 
ultralight axions, a prime DM candidate. Thus the scalaron in the Hu-Sawicki 
model of $\mathit{f(R)}$ gravity behaves as DM. Further, a study on the 
evolution of the scalaron mass with the redshift is also carried out, which 
depicts that scalaron becomes light with expansion of the Universe and with 
different rates at different stages of the Universe.
\end{abstract}


\maketitle
\section{Introduction}

One of the major challenges of present day physics and cosmology is 
uncovering the nature of mysterious dark matter (DM) 
\cite{bertone, swart, tim, frenk, strigari}. The first prediction of the 
existence of DM dates back to the early 1930s when J.\ H.\ Oort first 
postulated that 
more mass must be contained within the Milky Way galaxy over and above the 
visible amount to hold the stars in their respective orbits \cite{oort}. Then 
came Swiss astronomer Fritz Zwicky, who studied the Coma Cluster and found 
similar evidence of missing mass within the cluster \cite{zwicky, Zwicky}. 
Two most prominent pieces of evidences that argue about the existence of DM 
are namely, 
galactic rotation curve \cite{rubin, young, borriello} and gravitational 
lensing \cite{young, massey}. Various other evidences that support for 
non-luminous matter existence include the large mass-to-light ratio in
galaxy clusters \cite{lewis}, very high x-ray luminosity of the Bullet Cluster
(1E0657--558) \cite{clowe}, density parameters obtained from the observations 
of distant type Ia supernovae \cite{reiss, perlmutter}, etc. The Planck 
satellite observations of the Cosmic Microwave Background (CMB) radiation 
\cite{planck} have also provided fascinating evidence for the existence of DM 
through the determination of the cosmological parameters. These observations 
have also confirmed the standard $\Lambda$Cold Dark Matter ($\Lambda$CDM) 
cosmological paradigm by showing that baryonic matter alone cannot explain the 
mass content of the Universe. Observations of the Bullet Cluster also provide 
strong evidence for the existence of DM. In this cluster, the baryonic matter 
and the DM components are separated due to a long-ago collision of its two 
components \cite{clowe, Massey}. Using the Planck data \cite{planck} on the CMB 
radiation, measurements of the cosmological parameters imply that the Universe 
is composed of $\sim 4-5\%$ baryons, $\sim 25\%$ non-baryonic DM, and 
$\sim 70\%$ dark energy. A few particles have been claimed as DM candidates 
\cite{Bertone, garrett, feng}. Some of them are, namely, weakly interacting 
massive particles (WIMPs), standard model (SM) neutrinos, sterile neutrinos, 
axions, supersymmetric candidates (neutralinos, sneutrinos, gravitinos, 
axinos), etc. WIMPs are non-baryonic, which includes lightest SUSY particles, 
specially the neutralino, and it is considered as the most probable candidate 
of DM. Axions, which are also prime candidates of DM, are bosons that were 
first proposed to solve the strong CP problem. However, after almost eight 
decades since the DM concept was introduced, the DM particle is still missing 
from the table of elementary particles of nature, i.e.\ the fundamental nature 
of DM remains a mystery.

In the last three decades various issues and consequent limitations (specially,
the limitation related to the late time cosmic acceleration 
\cite{reiss, perlmutter}) of Einstein's General Theory of Relativity (GTR) 
have come to light, leading 
to the conclusion that GTR is not the ultimate theory of gravitational 
interaction. Theories of modified gravity \cite{capozziello} were proposed so 
that the gravitational interaction other than the one described by GTR, could 
be justified. In modified gravity models, we have $\mathit{f(R)}$ gravity 
models, Braneworld models, Gauss-Bonnet dark energy models, 
etc.\ \cite{clifton, oikonomou, odintsov, sergei}. Out of these, 
the simplest class of modified gravity theories is the $\mathit{f(R)}$ 
gravity \cite{faraoni, felice}. Here, modification occurs in the part 
describing 
the geometry of Einstein's field equation. It is modified by replacing the 
Ricci scalar $R$ of the Einstein-Hilbert action with a function $\mathit{f(R)}$ 
of $R$. There are two variational approaches to derive the field equations 
in $\mathit{f(R)}$ models, {\it {(i)}} metric formalism and {\it {(ii)}} 
palatini formalism. In metric formalism, matter is minimally coupled with 
the metric, and the energy-momentum tensor is independently conserved. 
In Palatini formalism, the metric as well as the connection are taken as 
independent variables. Here, the Riemann tensor as well as the Ricci tensor 
are constructed with the independent connection.

As the fundamental nature of DM is still a mystery, various theoretical as well
as observational studies have been carried out to understand DM. A plethora of 
research works have been devoted to explain the effects of DM in alternative 
theories of gravity \cite{katsuragawa, cembranos, yadav, harko, sen, riazi}. 
The formation of large-scale structure in the Universe dominated by DM and 
driven 
to accelerated expansion by $\mathit{f(R)}$ gravity in the Palatini formalism 
is studied in Ref.\ \cite{koivisto}. Also, DM and dark energy have been 
studied with scalar fields in Ref.\ \cite{varun}. The dynamics of the scalar 
fields in $\mathit{f(R)}$ gravity have also been studied in Ref.\ 
\cite{goswami}. These studies have motivated us to take into consideration of 
modified theories of gravity to understand DM. In our work, we discuss the DM 
problem in $\mathit{f(R)}$ gravity using its scalar degree of freedom. We apply
the metric formalism in $\mathit{f(R)}$ gravity to explain DM. The scalar field
plays an important role in $\mathit{f(R)}$ gravity. This scalar field also 
called scalaron is derived from the modification of gravitational theory. We 
particularly consider the Hu-Sawicki model \cite{sawicki}, proposed by Wayne Hu
and Ignacy Sawicki (2007), to explain the effects of DM. We have chosen this 
$\mathit{f(R)}$ gravity model \cite{shinichi} as it is one of the few known 
viable models within the framework of modified gravity which is able to 
satisfy solar system tests.

Our paper is organized in six sections. In the section \ref{sec.2}, we discuss 
the simplest type of modified gravity, i.e.\ $\mathit{f(R)}$ gravity using 
metric formalism. Here, we obtain the modified field equations. The Weyl 
transformation from the Jordan frame to the Einstein frame adds an extra 
degree of freedom. Then we obtain the potential of the scalaron which helps us 
to calculate the effective potential of the scalaron. In the section 
\ref{sec.3}, we obtain the mass of the scalaron as a function of matter 
effect $T_\mu^\mu$ . In section \ref{sec.4}, the chameleon mechanism is 
studied in the framework of the Hu-Sawicki model of $\mathit{f(R)}$ gravity. 
Further in section \ref{sec.5}, we discuss the properties of the scalaron in 
the present Universe, and the evolution of the scalaron mass with 
expanding Universe. Finally, in section \ref{sec.6}, we conclude and discuss 
the results of our work.

\section{$\mathbf{f(R)}$ gravity and conformal transformation}
\label{sec.2}

\subsection{Field Equations}

We consider the modification of the Einstein-Hilbert action \cite{starobinsky} 
as given by
\begin{equation}
S = \frac{1}{2\kappa^2}\int d^4x \sqrt{-g}\,(R + f(R)) + \int d^4x \sqrt{-g}\,\mathcal{L}_m\!\left[g^{\mu\nu}, \Phi\right],
\label{eqn.1}
\end{equation}
where $\mathit{f(R)}$ is a function of the Ricci scalar $R$ and 
$\kappa^2 = 8\pi G = 1/M_{pl}^2$ with $\hbar = c = 1$. $M_{pl}$ is the 
(reduced) Planck Mass $\sim 2 \times 10^{18}$ GeV. $\mathcal{L}_m$ is the 
Lagrangian of the matter part of the action with matter field $\Phi$.
Here, we apply the metric formalism. So, variation of the action \eqref{eqn.1} 
with respect to the metric $g_{\mu\nu}$ leads to the equation of motion,
\begin{equation}
\left(R_{\mu\nu} - \frac{1}{2}R\,g_{\mu\nu}\right) + f_R(R)R_{\mu\nu} - g_{\mu\nu}\! \left[\frac{f(R)}{2} - \square f_R(R)\right] - \nabla_\mu \nabla_\nu f_R(R) = \kappa^2 T_{\mu\nu},
\label{eqn.2}
\end{equation}
where $\mathit{f_R(R)}$ is the derivative of $\mathit{f(R)}$ with respect to 
$R$. And, the energy momentum tensor $T_{\mu\nu}$ is given by
\begin{equation}
T_{\mu\nu}\big[g^{\mu\nu}, \Phi\big] = \frac{-\,2}{\sqrt{-g}}\, \frac{\delta\left(\sqrt{-g}\,\mathcal{L}_m\!\left[g^{\mu\nu}, \Phi\right]\right)}{\delta g_{\mu\nu}}.
\label{eqn.3}
\end{equation}
The term within the parentheses in Eq.\ (\ref{eqn.2}) is the Einstein tensor
and hence remaining terms together on the left side give the modified part of 
the Einstein's field equation.
 
Now, the trace of Eq.\ \eqref{eqn.2} is given as
\begin{equation}
3\,\square f_R(R) - R + Rf_R(R) - 2f(R) = \kappa^2 T_\mu^\mu,
\label{eqn.4}
\end{equation}
which shows that the Ricci scalar $R$ becomes dynamical to be present as a new 
scalar degree of freedom if $f(R) \neq R$. In GTR, where 
$f(R) = R$, Eq.\ \eqref{eqn.4} simply leads to the trivial solution 
$R = -\,\frac{1}{2}\kappa^2\, T_\mu^\mu$. In a useful form this equation can
be written as
\begin{equation}
\square f_R(R) =  \frac{1}{3}\!\left[R + 2f(R) - Rf_R(R) + \kappa^2 T_\mu^\mu\right].
\label{eqn.4a}
\end{equation} 
This equation is analogous to the scalar field equation of Klein-Gordon type
in a potential if we consider that $f_R(R)$ is the extra scalar degree of 
freedom of $f(R)$ gravity. 

\subsection{Weyl Transformation}

To study the dynamics of the new scalar field, we require the Weyl 
transformation. It is the frame transformation from the Jordan frame 
$g_{\mu\nu}$ to Einstein frame $\tilde g_{\mu\nu}$ \cite{Katsuragawa}, which 
we consider in the form:
\begin{equation}
g_{\mu\nu}\rightarrow \tilde g_{\mu\nu} = e^{\sqrt{\frac{2}{3}}\kappa\phi}g_{\mu\nu} \equiv \left(1 + f_R(R)\right)g_{\mu\nu},
\label{eqn.5}
\end{equation}
where $\phi$ is the new scalar field named as scalaron. It is to be
noted that via this transformation the Ricci scalar $R$ can be expressed as
a function of the scalaron field $\phi$. Our intention is to 
obtain the field equations in the Einstein frame to study the effect of the 
scalaron as dark matter. For this purpose we have to write the action 
\eqref{eqn.1} in the Einstein frame using this transformation Eq.\ 
\eqref{eqn.5}. However, before doing so, let us rewrite the action 
\eqref{eqn.1} as given by
\begin{equation}
S = \int d^4x \sqrt{-g} \left[\frac{R + Rf_R(R)}{2\kappa^2} - U(R) \right] + \int d^4x \sqrt{-g}\,\mathcal{L}_m\!\left[g^{\mu\nu}, \Phi\right],
\label{eqn.6}
\end{equation}
where
\begin{equation}
U(R) = \frac{1}{2\kappa^2}\big[Rf_R(R) - f(R)\big],
\label{eqn.7}
\end{equation}
which can be treated as the potential of a new supplementary scalar field in 
the scalar-tensor equivalence of $f(R)$ gravity \cite{Velasquez, ruf} in Jordan 
frame itself \cite{Velasquez}. This means that there is an extra scalar degree 
of freedom of $f(R)$ gravity in Jordan frame as already specified by the trace 
Eq.\ \eqref{eqn.4}. Later we will see that $U(R)$ is related with the scalar 
field potential in the Einstein frame.    

Now, under the Weyl transformation \eqref{eqn.5}, the Ricci scalar $R$ in
Jordan frame and $\tilde{R}$ in Einstein frame is related by the equation:
\begin{equation}
R = \Omega^2 \left[\tilde{R} + 6\,\square \ln\Omega - 6\, \tilde g^{\,\mu\nu} \nabla_\mu (\ln\Omega) \nabla_\nu (\ln\Omega)\right],
\label{eqn.8}
\end{equation}
where $\Omega^2 \equiv 1 + f_R(R)$ is the conformal factor.
Thus, using the transformation \eqref{eqn.5} and the consequent relation 
\eqref{eqn.8}, the original action \eqref{eqn.1} can be written in Einstein
frame as
\begin{equation}
S = \frac{1}{2\kappa^2}\!\int d^4x \sqrt{-\tilde{g}}\tilde{R} + \int d^4x \sqrt{-\tilde{g}}\left[-\,\frac{1}{2}\,\tilde g^{\mu\nu}(\partial_\mu\phi)(\partial_\nu\phi) - V(\phi)\right] + \int d^4x \sqrt{-\tilde{g}}\, e^{-2\sqrt{\frac{2}{3}}\kappa\phi}\,\mathcal{L}_m\!\left[g^{\mu\nu}, \Phi\right],
\label{eqn.9}
\end{equation}
where
\begin{equation}
V(\phi) = \frac{U(\phi)}{\Omega^2} = \frac{1}{2\kappa^2}\,\frac{Rf_R(R) - f(R)}{(1 + f_R(R))^2}
\label{eqn.10}
\end{equation}
is the potential of the scalaron field $\phi$. Since, here $R = R(\phi)$, so in
view of this we may consider that $U(R) \equiv U(\phi)$. This also justifies our
conjecture mentioned above related with it in Eq.\ \eqref{eqn.7}. Hence, as usual 
Eq.\ \eqref{eqn.10} also shows the relationship of the scalar field 
potentials in Jordan frame and Einstein frame. 

The equation of motion for the scalaron field is obtained by the variation of 
Eq.\ \eqref{eqn.9} with respect to the scalaron field $\phi$, which is given 
as
\begin{equation}
\sqrt{-\tilde{g}}\left[\tilde{\square}\phi - V(\phi),_\phi\right] + \frac{\delta}{\delta\phi}\left(\sqrt{-g}\,\mathcal{L}_m\!\left[g^{\mu\nu}, \Phi\right]\right) = 0,
\label{eqn.11}
\end{equation}
where
\begin{equation}
\tilde{\square} = \frac{1}{\sqrt{-\tilde{g}}}\,\partial_\mu(\sqrt{-\tilde{g}}\,\tilde{g}^{\,\mu\nu}\partial_\nu).
\label{eqn.12}
\end{equation}
Since,
\begin{equation}
\delta g^{\mu\nu} = \frac{2\kappa}{\sqrt{6}}\,e^{\sqrt{\frac{2}{3}}\kappa\phi}\,\delta\phi\, \tilde{g}^{\,\mu\nu} = \frac{2\kappa}{\sqrt{6}}\, g^{\mu\nu} \delta\phi.
\label{eqn.13}
\end{equation}
So, we can have
\begin{equation}
\frac{\delta}{\delta\phi} = \frac{2\kappa}{\sqrt{6}}\, g^{\mu\nu} \frac{\delta}{\delta g^{\mu\nu}}.
\label{eqn.14}
\end{equation}
Substituting Eq.\ \eqref{eqn.14} in Eq.\ \eqref{eqn.11}, we get
\begin{equation}
\tilde{\square}\phi = V(\phi),_\phi\; +\; \frac{\kappa}{\sqrt{6}}\, e^{-2\sqrt{\frac{2}{3}}\kappa\phi}\, T_\mu^\mu.
\label{eqn.15}
\end{equation}
This equation is similar to the Klein-Gordon equation for a scalar field with 
an effective potential and it corresponds to Eq.\ \eqref{eqn.4a} in Jordan
frame as mentioned above. Thus, we may rewrite this equation as
\begin{equation}
\tilde{\square}\phi = \frac{d V_{eff}(\phi)}{d\phi},
\label{eqn.15a}
\end{equation}
where
\begin{equation}
\frac{d V_{eff}(\phi)}{d\phi} = V(\phi),_\phi\; +\; \frac{\kappa}{\sqrt{6}}\, e^{-2\sqrt{\frac{2}{3}}\kappa\phi}\, T_\mu^\mu.
\label{eqn.15b}
\end{equation}
Integrating this equation, we obtain the effective potential of the scalaron as
\begin{equation}
V_{eff}(\phi) = V(\phi) - \frac{1}{4}\, e^{-2\sqrt{\frac{2}{3}}\kappa\phi}\, T_\mu^\mu.
\label{eqn.16}
\end{equation}
It is to be noted that the effective potential of scalaron includes the trace 
of the energy-momentum tensor $T_\mu^\mu$, which indicates that the matter 
distribution has an important role to play on the potential of the scalaron. 
It turns out
that the effective mass of scalaron depends on the type of matter distribution 
and this is a necessary requirement for scalaron to exhibit the chameleon 
mechanism, which will be clear from the next sections.

\section{Scalaron Mass}
\label{sec.3}

Here, we will derive the mass of scalaron taking into account the effect of 
the matter distribution $T_\mu^\mu$ from the extremum condition of the 
effective potential. For this purpose we first calculate the derivative of 
the effective potential as 
\begin{equation}
\begin{split}
V_{eff}(\phi),_\phi & = V(\phi),_\phi\; +\; \frac{\kappa}{\sqrt{6}}\, e^{-2\sqrt{\frac{2}{3}}\kappa\phi}\, T_\mu^\mu \\[5pt]
& = \frac{1}{\sqrt{6}\kappa}\!\left[\frac{R(1 - f_R(R)) + 2f(R) + \kappa^2 T_\mu^\mu}{(1 + f_R(R))^2}\right].
\end{split}
\label{eqn.17}
\end{equation}
It would be worthwhile to mention that the comparison of this equation
with Eq.\ \eqref{eqn.4a} justifies our earlier conclusion about the 
later equation in relation with Eq.\ \eqref{eqn.15}. This, in fact, is the
result of the correspondence of physical situations between the Jordan frame 
and the Einstein frame.
  
The effective potential has an extremum at $\phi\,=\, \phi_{\text{min}}$. So 
at this situation,
\begin{equation}
V_{eff}(\phi),_{\phi} \big|_{\phi\, =\, \phi_{\text{min}}} = 0.
\label{eqn.18}
\end{equation}
Applying this condition to Eq.\ \eqref{eqn.17} through the Weyl 
transformation, we get,
\begin{equation}
R_0 - R_0f_R(R_0) + 2f(R_0) + \kappa^2 T_\mu^\mu = 0,
\label{eqn.19}
\end{equation}
where $R_0$ is the value of $R$ corresponding to $\phi\,=\, \phi_{\text{min}}$.
This equation can be used to find the value of $R_0$ for a given matter 
distribution.
Now, the second derivative of the effective potential is obtained as
\begin{equation}
\begin{split}
V_{eff}(\phi),_{\phi\phi} & = V(\phi),_{\phi\phi} \;-\; \frac{2\kappa^2}{3}\,e^{-2\sqrt{\frac{2}{3}}\kappa\phi}\, T_\mu^\mu \\[5pt]
& = \frac{Rf_R(R) - 3R - 4f(R)}{3(1 + f_R(R))^2} + \frac{1}{3f_{RR}(R)} - \frac{2}{3}\frac{\kappa^2 T_\mu^\mu}{(1 + f_R(R))^2}.
\end{split}
\label{eqn.20}
\end{equation}
At extremum, the second derivative of the effective potential gives us the 
square of the scalaron mass. Thus using Eq.\ \eqref{eqn.19} in this Eq.\ 
\eqref{eqn.20} the scalaron mass square can be obtained as follows:
\begin{equation}
\begin{split}
m_\phi^2 & = V_{eff}(\phi),_{\phi\phi}\big|_{\phi\, =\, \phi_{\text{min}}} \\[5pt]
& = \frac{1}{3(1 + f_R(R_0))}\left[\frac{1 + f_R(R_0)}{f_{RR}(R_0)} - R_0\right].
\label{eqn.21}
\end{split}
\end{equation}
In the high curvature regime, where $|f_R(R) \ll 1|$ and $|f/R| \ll 1$, this
mass square term can be written as
\begin{equation}
m_\phi^2 \approx \frac{1}{3}\!\left[\frac{1 + f_R(R_0)}{f_{RR}(R_0)} - R_0\right].
\label{eqn.22}
\end{equation}
Although this equation looks independent of matter distribution, i.e.\ of 
$T_\mu^\mu$, but as we have already seen that $R_0$ is determined by 
Eq.\ \eqref{eqn.19} as a function of the trace of the energy-momentum tensor 
$T_\mu^\mu$. Hence, the scalaron mass changes according to the trace of the 
energy-momentum tensor $T_\mu^\mu$, i.e. with the matter distributions.

\section{Hu-Sawicki $\mathbf{f(R)}$ gravity and chameleon mechanism}
\label{sec.4}

To study the chameleon mechanism in the behaviour of scalaron field, we will
use the Hu-Sawicki $\mathit{f(R)}$ gravity model. The Hu-Sawicki model of 
$\mathit{f(R)}$ gravity was proposed by Wayne Hu and 
Ignacy Sawicki in 2007 \cite{sawicki}. It represents one of the few known 
viable functional forms of $\mathit{f(R)}$ gravity with the interesting 
feature of being able to satisfy solar system tests. This model is introduced 
to explain the current situation of the accelerating Universe without taking 
into account a cosmological constant term. In this model,

\begin{equation}
f(R) = -\,m^2 \frac{c_1(R/m^2)^n}{c_2(R/m^2)^n + 1},
\label{eqn.23}
\end{equation}
where $n > 0$, $c_1$ and $c_2$ are two dimensionless model parameters. The 
parameter $m^2$ represents the mass (energy) scale and for our case we will
consider it equivalent to the present energy scale of the Universe, which is 
the current observed value of the cosmological constant $\Lambda$ \cite{Planck}.   
%\begin{equation}
%m^2 \equiv \frac{\kappa^2 \bar{\rho}_o}{3} = (8315\; \text{Mpc})^{-2}\left(\frac{\Omega_m h^2}{0.13}\right)
%\label{eqn.24}
%\end{equation}
%with $\bar{\rho}_o$ as the average matter density of the present Universe.
In the high curvature regime, where $R \gg m^2$, Eq.\ \eqref{eqn.23} can be 
simplified as
\begin{equation}
\lim_{m^2/R \to 0} f(R) \approx -\, \frac{c_1}{c_2}\,m^2 + \frac{c_1}{c_2^2}\,m^2 \left(\frac{m^2}{R}\right)^n\!\!\!.
\label{eqn.25}
\end{equation}
Therefore, we can have
\begin{equation}
f_R(R) = -\,\frac{c_1}{c_2^2}\,n\! \left(\frac{m^2}{R}\right)^{n + 1}
\label{eqn.26}
\end{equation}
and
\begin{equation}
f_{RR}(R) = \frac{n(n + 1)}{R}\, \frac{c_1}{c_2^2}\left(\frac{m^2}{R}\right)^{n + 1}\!\!\!\!\!\!\!.
\label{eqn.27}
\end{equation}

Substituting Eqs.\ \eqref{eqn.26} and \eqref{eqn.27} into Eq.\ \eqref{eqn.22} 
for $R = R_0$ we may obtain the expression for the mass square of 
scalaron in the Hu-Sawicki model of $\mathit{f(R)}$ gravity, which is given by
\begin{equation}
m_\phi^2 = \frac{1}{3}\,\frac{R_0\!\left[1 - n(n + 2)\, \frac{c_1}{c_2^2}\! \left(\frac{m^2}{R_0}\right)^{n + 1}\right]}{n(n + 1)\, \frac{c_1}{c_2^2}\! \left(\frac{m^2}{R_0}\right)^{n + 1}}.
\label{eqn.28}
\end{equation}

\begin{figure}[!h]
\centerline{
\includegraphics[scale = 0.35]{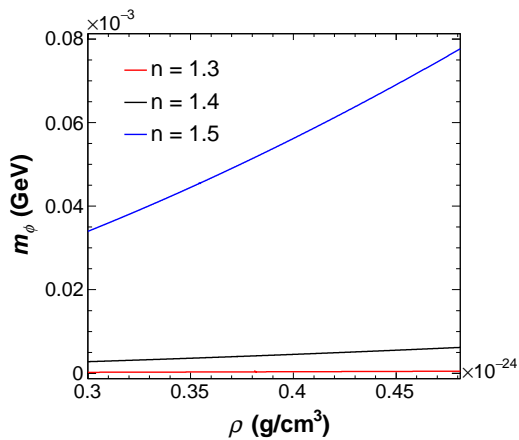}}
\caption{Variations of scalaron mass $m_\phi$ with energy density $\rho$ for
the model parameters $c_1 = 2.23$ and $c_2 = 0.08$ \cite{santos}, but for three 
different values of the parameter $n$. This set of values of $c_1$ and $c_2$
are used in all other figures wherever necessary.}
\label{fig.1}
\end{figure}

\noindent Similarly, substituting these equations into Eq.\ \eqref{eqn.19}, 
the equation to determine $R_0$ from the condition of minimum of the potential 
for this model is found as
\begin{equation}
R_0 + (n - 2)\, \frac{c_1}{c_2^2}\, m^2\! \left(\frac{m^2}{R_0}\right)^n -\, 2\, \frac{c_1}{c_2}\,m^2 + \kappa^2 T_\mu^\mu = 0.
\label{eqn.29}
\end{equation}
In high-curvature regime $R \gg m^2$, the above equation gives,
\begin{equation}
R_0 \approx -\, \kappa^2 T_\mu^\mu + 2\, \frac{c_1}{c_2}\,m^2.
\label{eqn.30}
\end{equation}
If we assume the pressureless ($p = 0$) dust model of the Universe, then the 
matter contribution is approximately expressed as $T_\mu^\mu = -\,\rho$, where 
$\rho$ is the matter-energy density of the Universe. Hence, for this model of 
the Universe Eq.\ \eqref{eqn.30} takes the form:
\begin{equation}
R_0 \approx \kappa^2 \rho + 2\, \frac{c_1}{c_2}\,m^2.
\label{eqn.31}
\end{equation}
Again Eq.\ \eqref{eqn.28} can be rewritten as
\begin{equation}
m_\phi^2 = \frac{m^2 c_2^2}{3n(n + 1)c_1}\left(\frac{R_0}{m^2}\right)^{n + 2} -\; \frac{(n + 2)R_0}{3(n + 1)}.
\label{eqn.32}
\end{equation}
As we already know that in high-curvature regime, $R/m^2 \gg 1$, so in this
case the second term on the R.H.S.\ of above equation becomes negligible
in comparison of the first, and hence for this case we may write, 
\begin{equation}
m_\phi^2 \approx \frac{m^2 c_2^2}{3n(n + 1)c_1}\left(\frac{R_0}{m^2}\right)^{n + 2}\!\!\!\!\!\!\!\!.
\label{eqn.33}
\end{equation}
Finally, substituting the value of $R_0$ obtained from Eq.\ \eqref{eqn.31} in 
Eq.\ \eqref{eqn.33}, we get the expression of the scalaron mass square in 
terms of matter density of the Universe and the Hu-Sawicki model parameters as
given by  
\begin{equation}
m_\phi^2 = \frac{\left(\kappa^2 \rho + 2\, \frac{c_1}{c_2}\,m^2 \right)^{n + 2}}{3n(n + 1)\,\frac{c_1}{c_2^2}\, m^{2(n + 1)}}.
\label{eqn.34}
\end{equation}
%Moreover, we have applied \cite{sawicki},
%\begin{equation}
%\frac{c_1}{c_2} \approx \frac{6\tilde{\Omega}_\Lambda}{\tilde{\Omega}_m}
%\label{eqn.35} 
%\end{equation}
%so that, Eq.\ \eqref{eqn.34} is left with only two variable parameters, i.e.\
%$n$ and $c_2$. As mentioned in Ref.\ \cite{sawicki}, Eq.\ \eqref{eqn.35} is 
%set in this way to approximate the expansion history of $\Lambda$CDM with a 
%cosmological constant parameter $\tilde{\Omega}_\Lambda$ and matter 
%density parameter $\tilde{\Omega}_m$ with respect to a fixed critical 
%value, hence, to control how closely the model represents $\Lambda$CDM. 
%Eq.\ \eqref{eqn.35} is valid as long as the high curvature regime holds. For
%our case we have taken  $\tilde{\Omega}_\Lambda = 0.76$ and  
%$\tilde{\Omega}_m = 0.24$.
It should be noted at this point that the best-fit values of the model 
parameters $c_1$ and $c_2$ are obtained from various observational tests as 
discussed in the Ref.\ \cite{santos}. However, from our study we found that 
the values of $c_1$ and $c_2$ have to be $>0$, otherwise the results become 
undefined. Hence, for our rest of the study we will use only one particular 
set of best-fit values of $c_1$ and $c_2$, i.e.\ $c_1 = 2.23$ and $c_2 = 0.08$ 
obtained from the observational $\mathit{H(z)}$ test \cite{santos}.

From Eq.\ \eqref{eqn.34}, we can see that the mass of the scalaron behaves as 
a monotonically increasing function of the matter density $\rho$ as depicted 
in Fig.\ \ref{fig.1}. This means that the scalaron becomes heavy in 
high-density region, whereas it becomes light in the low-density region of 
matter. This particular behaviour exhibited by the scalaron field is one of 
the screening mechanisms, which is known as the chameleon mechanism.
In Fig.\ \ref{fig.1}, we have chosen three values of $n$ keeping the value of 
$c_1 = 2.23$ and $c_2 = 0.08$ according to our consideration as mentioned above
to get a comparative assessment of the way the scalaron mass changes with 
respect to the matter density $\rho$ for different values of the parameter 
$n$. It is seen that for each ascending value of $n$, the scalaron mass 
increases rapidly.

\subsection{Singularity Problem}
There is a singularity problem in the Hu-Sawicki model of $\mathit{f(R)}$ 
gravity. To understand this problem that arises in the model we first consider 
the scalaron potential without taking into account the matter contribution. The
scalaron potential is already obtained as an explicit function of the Ricci 
scalar $R$, which is given by Eq.\ \eqref{eqn.10}. Using the Weyl 
transformation \eqref{eqn.5}, we may obtain the scalaron 
potential and effective potential as an explicit function of the scalaron 
field $\phi$. Thus, the scalaron potential given by Eq.\ \eqref{eqn.10} can be 
expressed in terms of $\phi$ for the Hu-Sawicki model as
\begin{equation}
\frac{V(\phi)}{V_0} = \frac{\displaystyle -\,n\,\frac{c_1}{c_2}\, -\, n\, \frac{c_1}{c_2^2}\left(\frac{1 - e^{\sqrt{2/3}\kappa\phi}}{n\frac{c_1}{c_2^2}}\right)^{\frac{n}{n + 1}} +\; c_1 \left(\frac{1 - e^{\sqrt{2/3}\kappa\phi}}{n\,\frac{c_1}{c_2^2}}\right)^{-\frac{n}{n + 1}}}{\displaystyle e^{2\sqrt{2/3}\kappa\phi}\left[c_2 \left(\frac{1 - e^{\sqrt{2/3}\kappa\phi}}{n\,\frac{c_1}{c_2^2}}\right)^{-\frac{n}{n + 1}}  + 1\right]},
\label{eqn.36}
\end{equation}
where $V_0 = m^2/2\kappa^2$ is the normalization factor and the Weyl conformal 
transformation \eqref{eqn.5} gives the relation between $R$ and $\phi$ as
\begin{equation}
e^{\sqrt{2/3}\kappa\phi} = 1 - \frac{c_1}{c_2^2}\, n \left(\frac{m^2}{R}\right)^{n + 1}\!\!\!\!\!\!\!.
\label{eqn.37}
\end{equation}
In this relation we have used Eq.\ \eqref{eqn.26}. 
\begin{figure}[!h]
\centerline{
\includegraphics[scale = 0.35]{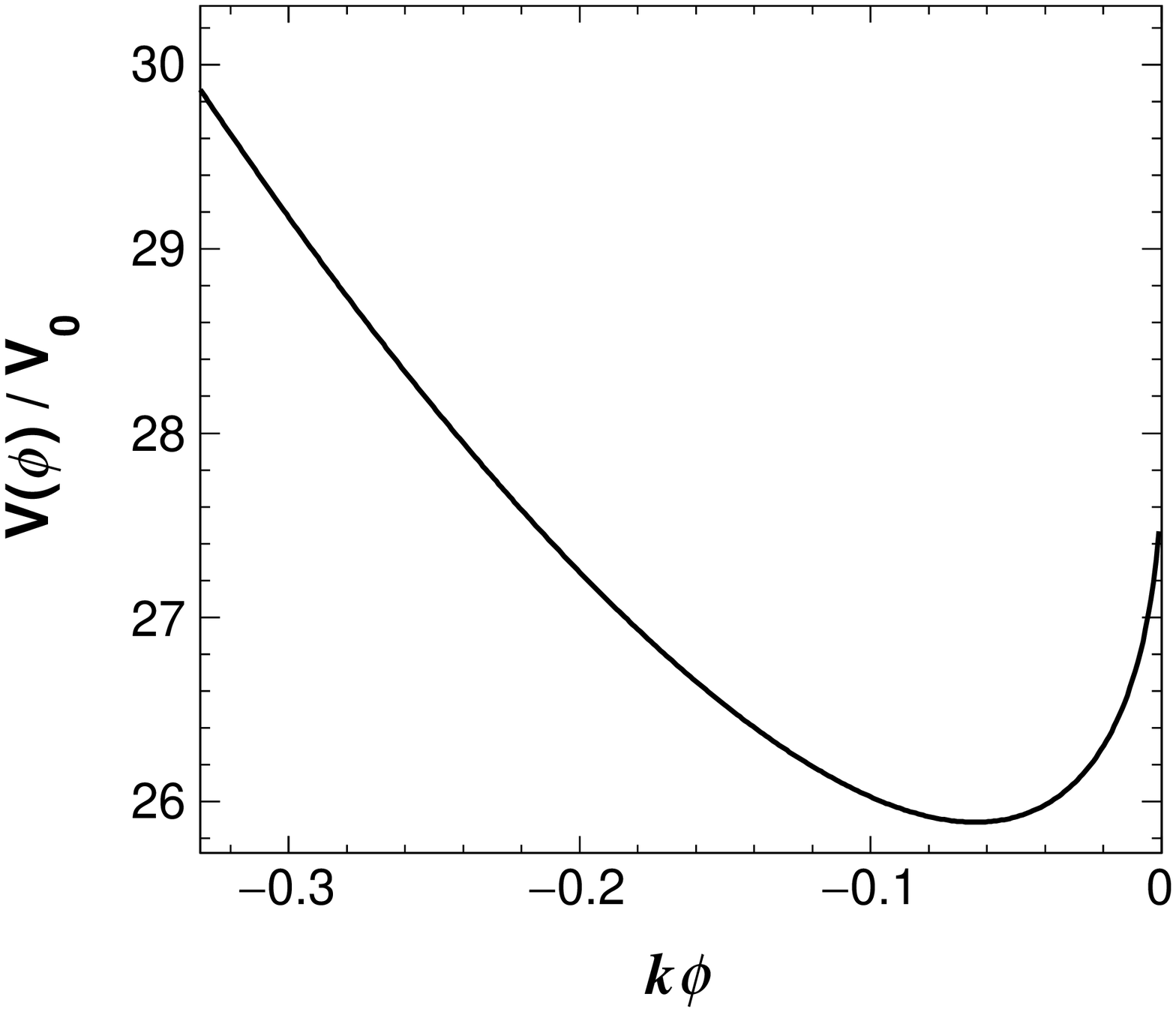} \hspace{0.5cm}
\includegraphics[scale = 0.35]{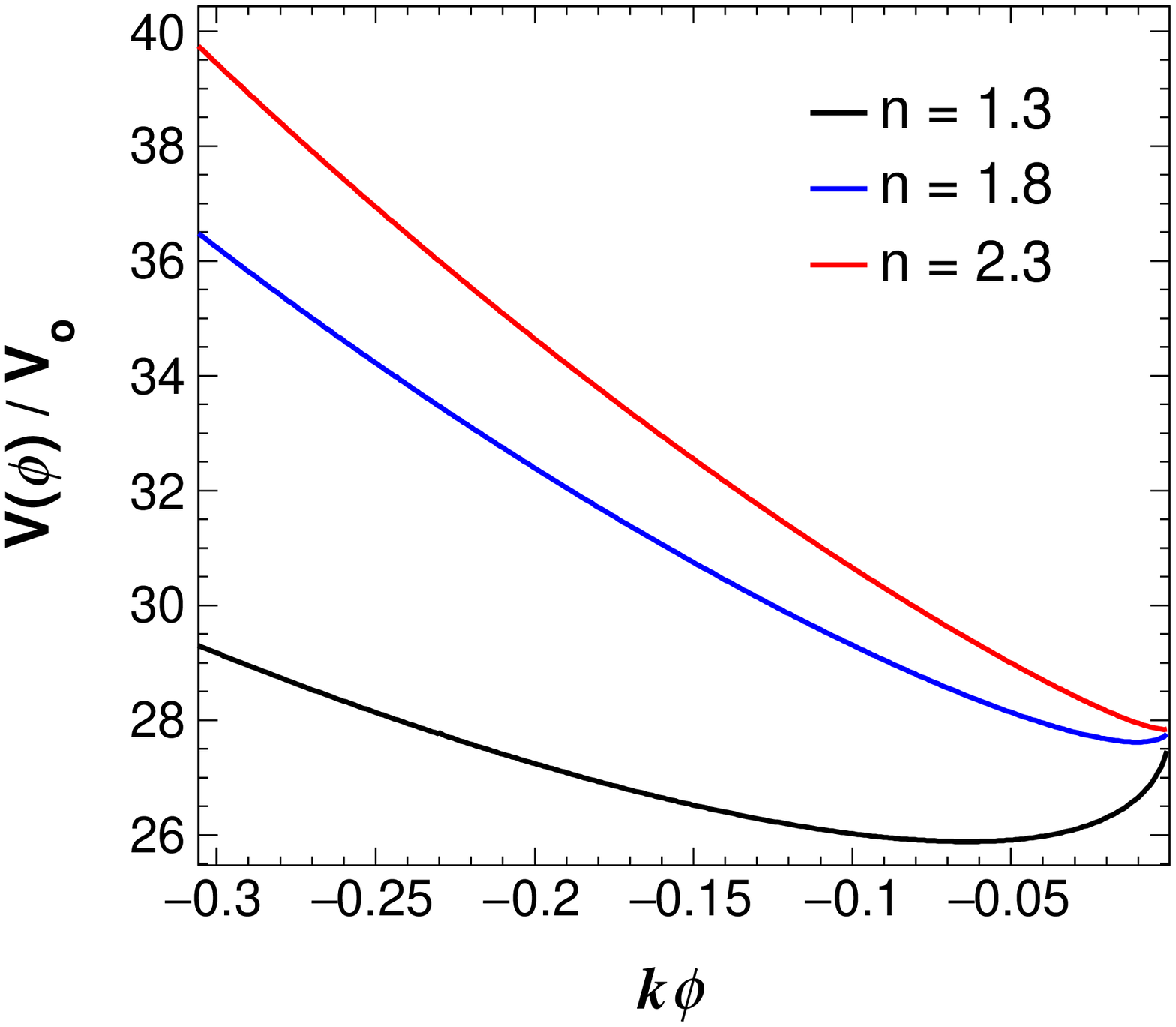}}
\caption{Variations of scalaron potential $V(\phi)$ for the Hu-Sawicki model 
as a function of scalaron field $\phi$. The left panel is for $n = 1.3$, while 
the right panel is for $n = 1.3, 1.8$ and $2.3$.}
\label{fig.2}
\end{figure}

\noindent The behaviour of scalaron potential as a function of the scalaron 
field for the Hu-Sawicki model given in the above Eq.\ \eqref{eqn.36} is 
depicted in Fig.\ \ref{fig.2}. In the first panel of this figure, the 
variation of the potential with the field is shown exclusively for the 
parameters chosen as $n = 1.3$, $c_1 = 2.23$ and $c_2 = 0.08$. As seen from 
this plot, the potential decreases moderately with the increasing values of 
$\kappa\phi$ from the negative side and becomes minimum at 
$\kappa\phi_{\text{min}} \sim -\,0.05$. After that the potential rises 
rapidly to terminate itself at $\kappa\phi = 0$ with a particular higher 
value. Again, we have studied the effect of variation of values of $n$ on 
the potential $V(\phi)$ for a fixed value of $c_1$ and $c_2$ as shown in the 
right panel of this figure. This panel is for $n = 1.3, 1.8, 2.3$ with values 
of $c_1$ and $c_2$ as in the case of the first panel. It is seen from this 
panel that as the parameter $n$ is varied to higher values, the scalaron 
potential increases rapidly from that for smaller values of $n$, but it 
descends to a particular value at $\kappa\phi = 0$ for all values of $n$. It 
is to be noted that in both of these two cases no potential exists for the 
positive values of the field. 

\begin{figure}[h]
\centerline{
\includegraphics[scale = 0.35]{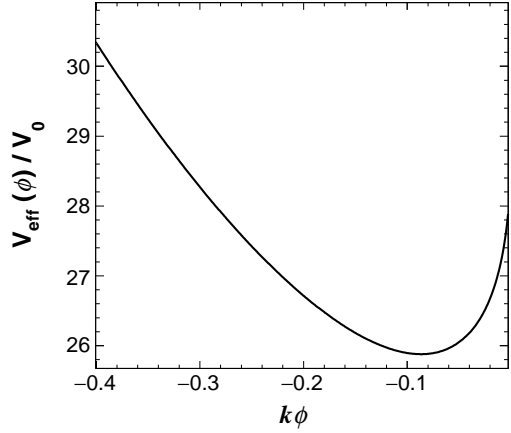}}
\caption{Variation of effective scalaron potential as a function of scalaron 
field for $-T_\mu^\mu \sim \rho_{\text{crit}}$. This potential is found to be 
independent of Hu-Sawaki model parameters $n$, $c_1$ and $c_2$.}
\label{fig.3}
\end{figure}

Next, we add the matter contribution term to study the effective potential of 
the scalaron field. From Eq.\ \eqref{eqn.16} the normalized effective potential
of the scalaron field is obtained as
\begin{equation}
\frac{V_{eff}(\phi)}{V_0} = \frac{V(\phi)}{V_0} - \frac{\kappa^2\, T_\mu^\mu}{2\,m^2\, e^{2\sqrt{2/3}\kappa\phi}}.
\label{eqn.38}
\end{equation}
This effective potential is calculated for the Hu-Sawicki model using the 
pressureless dust model of the Universe as mentioned earlier. The result of 
this calculation is depicted in Fig.\ \ref{fig.3}, which is independent of 
different model parameter values except $m^2$. Here, it is seen that by 
considering the matter contribution as a positive value of matter-energy 
density, the potential could be slightly shifted from its original minimum and 
also made to behave as an almost smoothly falling function of the field. Hence, 
the minimum of the effective potential occurs at closer to zero of the field 
value. We find that the minimum of the effective potential occurs so close to 
zero of the scalaron field that it can easily go to zero at the minimum. This 
indicates that the curvature singularity can be readily attained because 
$R \to \infty$ for $\phi \to 0$ as can be seen from Eq.\ \eqref{eqn.37}. This 
is the so-called curvature singularity problem \cite{frolov} and the infrared
modified viable $\mathit{f(R)}$ gravity models usually experience this problem.

\subsection{Higher Order Correction}

As can be seen from the previous subsection, the singularity problem arises 
in the Hu-Sawicki model in the large curvature regime. Hence, it is assumed 
that this problem can be mended by improving the potential structure of the 
scalaron field in the high curvature region. A well-known way to deviate the 
singularity problem is to add the higher curvature term \cite{nojiri}. In 
$\mathit{f(R)}$ gravity, the most suitable higher-curvature correction term 
is $\alpha R^2$, where $\alpha$ is a dimensionful constant parameter. With
this correction term the Hu-Sawicki model \eqref{eqn.23} can be modified as
\begin{equation}
f(R) = -\,m^2 \frac{c_1(R/m^2)^n}{c_2(R/m^2)^n + 1} + \alpha R^2.
\label{eqn.39}
\end{equation}
In the high curvature limit $R \gg m^2$, this Eq.\ \eqref{eqn.39} can be 
rewritten as
\begin{equation}
f(R) \approx -\, \frac{c_1}{c_2}\,m^2 + \frac{c_1}{c_2^2}\,m^2 \left(\frac{m^2}{R}\right)^n \!+\, \alpha R^2.
\label{eqn.40}
\end{equation}

\begin{figure}[h]
\centerline{
\includegraphics[scale = 0.35]{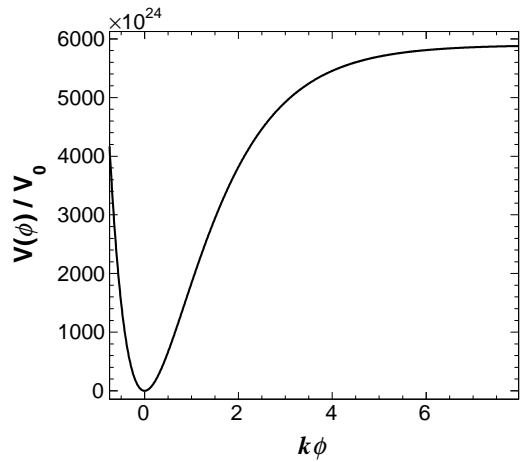}}
\caption{Variation of scalaron potential $V(\phi)$ as a function of 
scalaron field for the Hu-Sawicki model with the $R^2$ correction. The 
parameter $\alpha$ for this plot is taken as $\alpha = 10^{54}$ GeV$^{\,-2}$. 
This value of $\alpha$ is used for the subsequent figures if not mentioned
otherwise.}
\label{fig.4}
\end{figure}

\noindent Similarly, with this correction term in the large curvature regime, 
Eq.\ \eqref{eqn.37} which gives the relation between $R$ and $\phi$ can 
be modified as
\begin{equation}
e^{\sqrt{2/3}\kappa\phi} = 1 - \frac{c_1}{c_2^2}\, n \left(\frac{m^2}{R}\right)^{n + 1}\!\!\! +\, 2\alpha R.
\label{eqn.41}
\end{equation}
Therefore, for the $R^2$ corrected Hu-Sawicki model \eqref{eqn.39}, the scalaron 
potential is obtained as
\begin{equation}
\frac{V(\phi)}{V_0} = \frac{\displaystyle \frac{c_1}{c_2} + \left(e^{\sqrt{2/3}\kappa\phi} - 1\right)^2/4\,\alpha\, m^2}{e^{2\sqrt{2/3}\kappa\phi}}.
\label{eqn.42}
\end{equation}
The behaviour of this corrected potential $V(\phi)$ with respect to the 
scalaron field is shown in Fig.\ \ref{fig.4} for the parameters $c_1 = 2.23$, 
$c_2 = 0.08$ and $\alpha = 10^{54}$ GeV$^{\,-2}$. From this figure it is seen 
that for the positive value of $\phi$, the potential is modified to become 
finite in the large curvature region. The reason for choosing this particular
large value of $\alpha$ is mentioned below.

\begin{figure}[h]
\centerline{
\includegraphics[scale = 0.35]{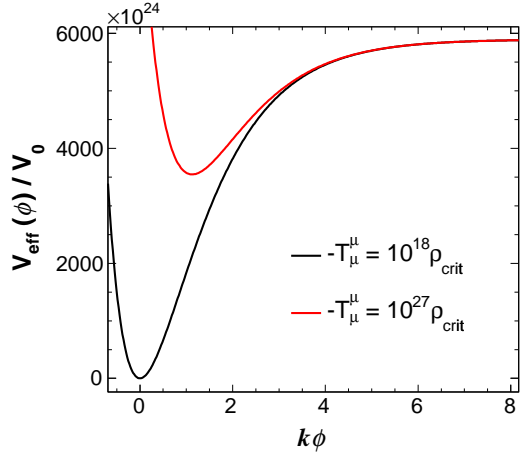}}
\caption{Variations of effective scalaron potential as a function of scalaron 
field for the Hu-Sawicki model with the $R^2$ correction are drawn for 
contributions of two matter distributions: $ - T_\mu^\mu \sim 10^{18} \rho_{\text{crit}}$ and $10^{27} \rho_{\text{crit}}$.}
\label{fig.5}
\end{figure}

\begin{figure}[h]
\centerline{
\includegraphics[scale=0.35]{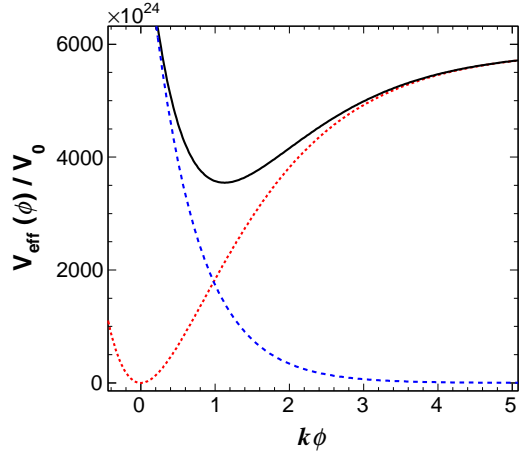}}
\caption{Variation of effective scalaron potential (black solid curve) as a 
function of scalaron field for Hu-Sawicki model with the $R^2$ correction is 
drawn with the contribution of matter distribution 
$ - T_\mu^\mu \sim 10^{27} \rho_{\text{crit}}$. The red dashed line shows 
the original potential shown in Fig.~\ref{fig.4} and the blue dashed line is 
for only the matter contribution with $ - T_\mu^\mu \sim 10^{27} \rho_{\text{crit}}$.}
\label{fig.6}
\end{figure}

Subsequently, we deduce the effective potential of the scalaron field for the 
$R^2$ corrected model \eqref{eqn.39}. As earlier, in this case also the effective potential is obtained in the form:
\begin{equation}
\frac{V_{eff}(\phi)}{V_0} = \frac{V(\phi)}{V_0} - \frac{\kappa^2\, T_\mu^\mu}{2\, m^2\, e^{2\sqrt{2/3}\kappa\phi}}.
\label{eqn.43}
\end{equation}
The variation of this effective potential as a function of $\kappa\phi$ is 
shown in Fig.~\ref{fig.5} for two matter distributions, viz., 
$ - T_\mu^\mu \sim 10^{18} \rho_{\text{crit}}$ and $10^{27} \rho_{\text{crit}}$.
 It is seen that with the increasing matter distribution to 
$10^{27} \rho_{\text{crit}}$ the minimum of the effective potential rises 
substantially from its approximately zero level for the matter distribution 
 $10^{18} \rho_{\text{crit}}$. It is also noticed that the matter distribution
$10^{18} \rho_{\text{crit}}$ has no significant effect on the potential as the
variation of the effective potential with this matter distribution is almost
similar to the plot of Fig.~\ref{fig.4}. Fig.~\ref{fig.6} is drawn to show 
explicitly the matter contribution to the effective potential \eqref{eqn.43}
with the matter distribution $10^{27} \rho_{\text{crit}}$. Moreover, we 
see from Fig.~\ref{fig.5} and \ref{fig.6} that $k\phi_{\text{min}} \sim 1$ is 
achieved when $- T_\mu^\mu \sim 10^{27} \rho_{\text{crit}}$. However, the 
value of $T_\mu^\mu$ for obtaining $k\phi_{\text{min}} \sim 1$ as well as the 
level of raising the minimum of the effective potential depends on the 
parameter values chosen. Although the value $\alpha = 10^{54}$ GeV$^{-2}$ is 
very large, this value of $\alpha$ is chosen so that the effect of matter 
contribution can be realised. Again, as we go on decreasing the value of 
$\alpha$, the value of $\kappa\phi_{\text{min}}$ also becomes smaller. As 
discussed in Ref.~\cite{Katsuragawa}, the background effect can be ignored if 
$\kappa\phi_{\text{min}}$ is small enough. Therefore, to obtain a smaller 
$k\phi_{\text{min}}$ as well as to account for the matter contribution we have 
to choose a reasonable value of $\alpha$. From the relation between $R$ and 
$\phi$ given by Eq.~\eqref{eqn.41} with $R^2$ correction it is seen that, 
here $R \to \infty$ corresponds to $\kappa\phi \to \infty$ 
(or $\phi \to \infty$) in contrast to the relation given by 
Eq.~\eqref{eqn.37}. This implies that here a small value of $k\phi$ corresponds 
to small curvature $R$. For example, when $k\phi_{\text{min}} = 0$ (as in the 
case of $- T_\mu^\mu \sim 10^{18} \rho_{\text{crit}}$), the corresponding 
value of $R$ is $R_0 = (c_1\,m^4/2\,\alpha\,c_2^2)^{1/3}$ for $n=1$, which is 
a finite value. Thus we see that for large matter-energy density 
in the high curvature regime, the potential has a minimum not at $R\to \infty$,
but at finite value of $R$ and also this minimum is not so minute as to let it
easily go to zero unlike the original effective potential shown in 
Fig.~\ref{fig.4}. Instead, here the singularity is moved away to infinity so 
that it is not easily attainable. Hence, the singularity problem is resolved.

At this stage, we want to calculate the mass of scalaron from the $R^2$ 
corrected Hu-Sawicki model \eqref{eqn.39}. In Fig.\ref{fig.1} we have already 
seen how the scalaron mass increases with matter energy density. This extremely
large mass of scalaron is a consequence of the singularity problem as 
mentioned above, because to subdue the singularity problem we need larger
energy density in the effective potential, which appears as some order higher 
in magnitude in the scalaron mass, as seen from Eq.~\eqref{eqn.34}. Thus
we should get a reasonably light scalaron in the high density region, once we 
solve the singularity problem. In this context, it needs to
be mentioned that according to the analysis in Ref.~\cite{katsuragawa}, the 
upper bound for the scalaron mass should be such that
\begin{equation}
m_\phi < \mathcal{O}(1)\;\text{GeV}.
\label{eqn.44}
\end{equation}
Now, from Eq.~\eqref{eqn.40}, we get
\begin{equation}
f_R(R) = -\,\frac{c_1}{c_2^2}\,n \left(\frac{m^2}{R}\right)^{n + 1} \!\!\!\!\!
+\, 2\alpha R.
\label{eqn.45}
\end{equation}
Similarly, from this equation, we obtain
\begin{equation}
f_{RR}(R) = n(n + 1)\, \frac{c_1}{c_2^2}\left(\frac{m^2}{R}\right)^{n + 1}\!\!\frac{1}{R} + 2\alpha.
\label{eqn.46}
\end{equation}
Substituting these values of $\mathit{f_R(R)}$ and $\mathit{f_{RR}(R)}$ for the
higher order correction in the Hu-Sawicki model in the scalaron
mass Eq.~\eqref{eqn.22}, we found the scalaron mass as
\begin{equation}
m_\phi^2 = \frac{1}{3}\! \left[ \frac{1 -n(n+2)\,\frac{c_1}{c_2^2} \left(\frac{m^2}{R_0}\right)^{n + 1}}{n(n + 1)\, \frac{c_1}{c_2^2}\,m^{-2}\left(\frac{m^2}{R_0}\right)^{n + 2} + 2\alpha}\right].
\label{eqn.47}
\end{equation}
Simplifying the above equation with the higher curvature condition 
$R \gg m^2$, the scalaron mass is deduced as
\begin{equation}
m_\phi^2 \approx \frac{1}{6\alpha}.
\label{eqn.48}
\end{equation}
Hence, the mass of the scalaron becomes constant and depends only on the value 
of $\alpha$ in the high curvature region.

\section{Properties of scalaron mass}
\label{sec.5}

In this section, we calculate the scalaron mass range relevant to the present 
Universe and discuss its implications. 
%We then review ways to obtain the correct relic abundance. 
Also we calculate and discuss the evolution of scalaron mass with the
cosmological redshift.

\subsection{Scalaron Mass Scale}
As already mentioned, if we consider our Universe under the dust 
approximation, then the typical energy density of galaxy can be 
found as $\rho_0 \sim 3-5 \times 10^{-25}$ g/cm$^3$ $\sim 2-3 \times 10^{-42}$ 
GeV$^4$ \cite{chiba}. The scalaron mass range corresponding to this range of 
energy density of galaxy can be computed from Eq.~\eqref{eqn.47} with a 
reasonable value of $\alpha = 10^{54}$ GeV$^{-2}$. Under these considerations 
the computed mass range of scalaron is found to be
\begin{equation}
m_\phi \sim 0.556 - 1.383 \times 10^{-23}\; \text{eV}.
\label{eqn.49}
\end{equation}

This result naturally satisfies the upper bound $m_\phi < \mathcal{O}(1)$ GeV, 
which means that the mass of the scalaron should be less than or of the order 
of $1$ GeV. The mass range of the scalaron calculated in Eq.~\eqref{eqn.49} 
is comparable to the mass of ultralight axions, which has an approximate mass 
of $\sim 10^{-22}$ eV \cite{joshua}. This DM candidate is referred to as the 
fuzzy cold dark matter (FCDM). In fact, the mass range given in Eq.~\eqref{eqn.49} is closer to the scalaron mass value $m_\phi \sim 3 - 5 \times 10^{-24}\; \text{eV}$ obtained for the Starobinsky model with $R^2$ correction
\cite{Katsuragawa}. As mentioned in \cite{hui}, this lower mass of the scalaron 
does not create any complication. Infact, it can be said that this scalaron 
makes small contributions to the amount of DM in the present Universe. However, 
DM candidates in the mass range $10^4\; \text{GeV} \gtrsim m_\phi \gtrsim 10^{-22}\;\text{eV}$ could potentially create huge amount of DM. From a theoretical 
point of view although such a low mass seems strange, yet observations as well 
as experimental evidences permit considerably lighter scalar fields 
($m \lesssim 10^{-33} \text{eV}$) \cite{barkana} so that observations of 
accelerated expansion of the Universe could be understood.
%Axions are one of the prime candidates of DM, the other prime candidate being 
%the weakly interacting massive particles (WIMPs). 
Furthermore, in Ref.\cite{nelson} the authors investigated the possibility
of a condensate of a very light vector boson as the dark matter and shows that 
the mass of such massive vector boson should satisfy the condition:
$m_\phi \geq \Omega_{DM}^2 H_0 \hbar = 6.6 \times 10^{-35}$ eV when the mass 
of the boson $\phi$ is comparable to the Hubble parameter $H$ for its upper
bound is equal to the Planck's mass. But for such a low mass the Compton 
wavelength is too large. In order to allow structure formation, the
Compton wavelength has to be on the kpc scale. This requirement gives a 
sharper bound on the lowest mass as
\begin{equation}
1 \text{kpc} < \frac{\hbar}{\Delta p} = \frac{\hbar}{m_\phi v_{esc}} \Rightarrow 
m_\phi \geq 1.67 \times 10^{-24}\; \text{eV}
\label{eqn.54}
\end{equation}
It is seen that our calculated scalaron mass range satisfies this lowest vector
boson mass condition. Usually, it is speculated that the mass of various 
DM candidates ranges from $10^{-22}$ eV (FCDM) \cite{joshua, salucci} to 
thousands of solar masses including the primordial black holes \cite{paul, bernard}. It would be appropriate to mention at this point that the relic 
abundance of scalaron as dark matter has been discussed in detail in 
Ref.\cite{Katsuragawa}. 

Moreover, it can be seen from the literature \cite{oikonomou20, odintsov20, 
oikonomou21} that when the scalaron mass $m_\phi$ approaches to Hubble 
parameter $H$, the scalaron field starts to oscillate, and the scalaron field 
oscillates in a slow-varying manner when $m_\phi \gg H$. Our work interprets 
this scalaron as an axion cold DM particle, which satisfies the chameleon 
mechanism. And for axions it has also been mentioned in the Ref.\ 
\cite{odintsov19} that during late times, at low curvature of the Universe, 
the slow-varying oscillation of the axion field leads to the fact that the 
axion energy density behaves as $\rho_\phi \sim a^{-3}$, where $a$ is the scale factor, and hence it satisfies the perfect fluid continuity equation. 
Whereas, in our work, the perfect fluid continuity equation is not satisfied. 
This is due to the fact that we have taken into account the pressureless dust 
model of the Universe. In this theory, the energy is not conserved and the 
energy density of the chameleon does not satisfy the continuity equation. Hence, with expansion of the Universe, the chameleon energy density does not behave 
as $\rho_\phi \sim a^{-3}$. 

Again, curiosity arises as to how the scalaron would behave near the surface 
as well as at the center of very dense objects, such as neutron stars. The 
density of neutron stars is around $10^{15} - 10^{16}$ g/cm$^3$ \cite{artyom}. 
Using this data in Eq.~\eqref{eqn.47}, we computed the scalaron mass inside a 
neutron star as well as near its surface, and found the scalaron mass 
$\sim 10^{-19}$ eV deep inside the star and $\sim 10^{-31}$ eV near the 
surface. Thus, the scalaron satisfies the chameleon mechanism, with lighter 
mass near the surface and heavier mass at the center of neutron stars. However,
detailed study is required in this direction, especially to see the 
characteristics of neutron stars in the light of chameleon mechanism.
Furthermore, the potential in our work is chameleonic and as such the 
inflationary constraints on the potential discussed in \cite{Oikonomou, 
Odintsov} do not apply. As our chameleonic potential is unaffected by 
inflationary constraints, the behaviour of the scalar field in our theory can 
be studied for neutron stars. Neutron stars have been studied in 
$\mathit{f(R)}$ theories that use the chameleon mechanism in \cite{babichev, brax}. In these works authors have demonstrated how the scalar field 
changes with the minimum of its effective potential at the core, near the 
boundary and outside the neutron stars.

\subsection{Evolution of Scalaron Mass}
Finally, in this section we study how the mass of the scalaron evolves as a 
function of cosmological redshift for specific parameter values of the model. 
This is to clearly see the behaviour of scalaron at different stages of the 
expanding Universe under the $\Lambda$CDM scenario. For this purpose, we used 
Eq.~\eqref{eqn.47} to obtain the required expression for the mass of the 
scalaron as a function of the cosmological redshift $z$. 

In the Hu-Sawicki model, for the flat $\Lambda$CDM model of cosmic expansion 
the Ricci scalar can be expressed as \cite{sawicki} 
\begin{equation}
R_0 \approx \kappa^2\bar{\rho}_0(1+z)^3 + 2\, \frac{c_1}{c_2}m^2,
\label{eqn.55}
\end{equation}
where $\bar{\rho}_0$ is the present average matter density of the Universe, 
since the matter density in the $\Lambda$CDM model is given by 
$\rho = \bar{\rho}_0(1 + z)^3$. Now, Eq.\ \eqref{eqn.55} can be used in 
Eq.~\eqref{eqn.47} to obtain the scalaron mass as a function of the redshift 
$z$.
%\begin{equation}
%m_{\phi}^2 = \frac{1}{3} \left[\frac{1 - n(n + 2)\frac{c_1}{c_2^2}\left(3(1 + z)^3 + 12\frac{\Omega_\Lambda}{\Omega_m}\right)^{-(n + 1)}}{n(n + 1)\frac{c_1}{c_2^2}m^{-2} \left(3(1 + z)^3 + 12\frac{\Omega_\Lambda}{\Omega_m}\right)^{-(n + 2)} + 2\alpha}\right],
%\label{eqn.57}
%\end{equation}
%where we have again used $\frac{c_1}{c_2} \approx \frac{6\Omega_\Lambda}{\Omega_m}$.

\begin{figure}[!h]
\centerline{
\includegraphics[scale = 0.28]{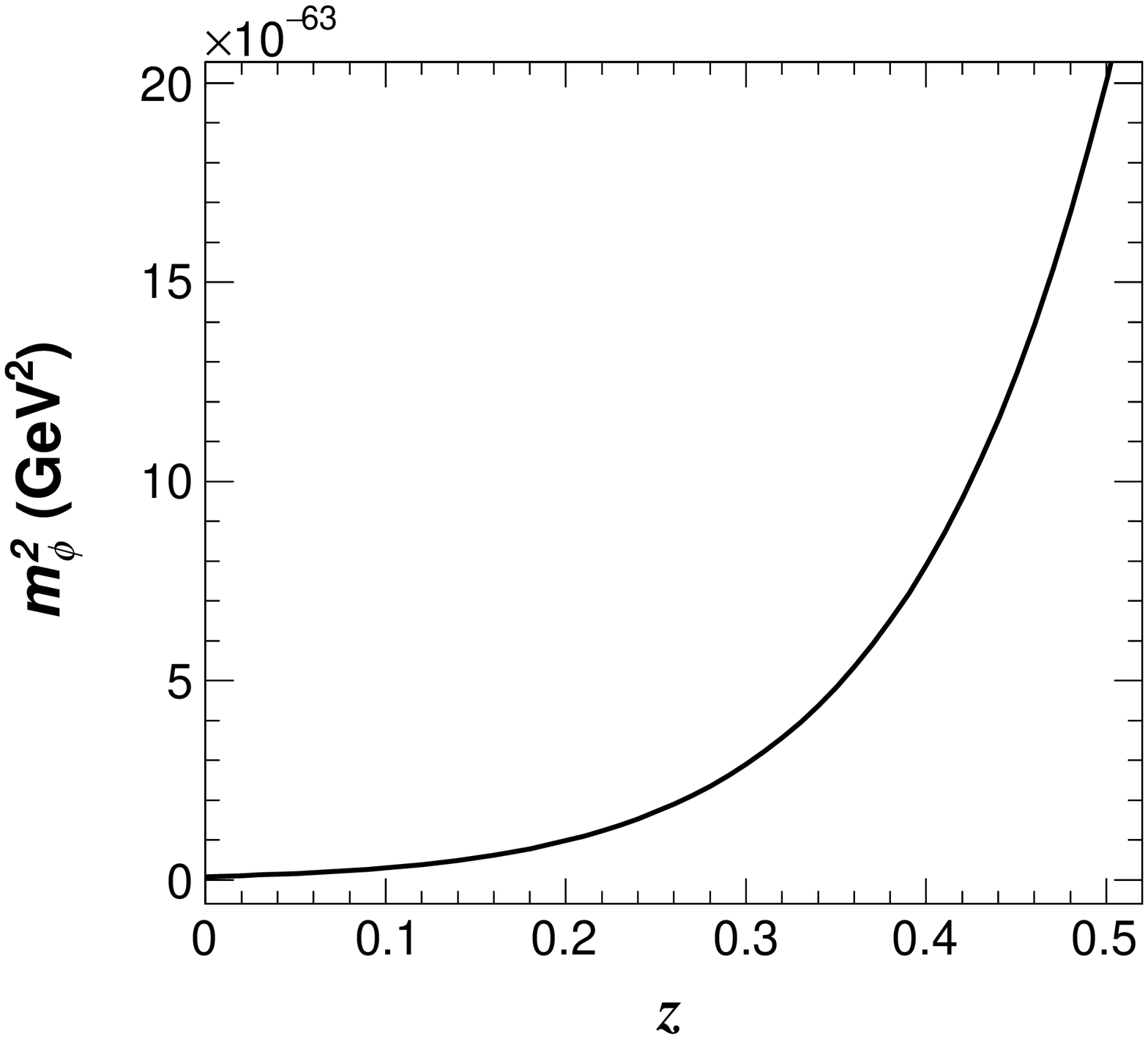} \hspace{0.2cm}
\includegraphics[scale = 0.28]{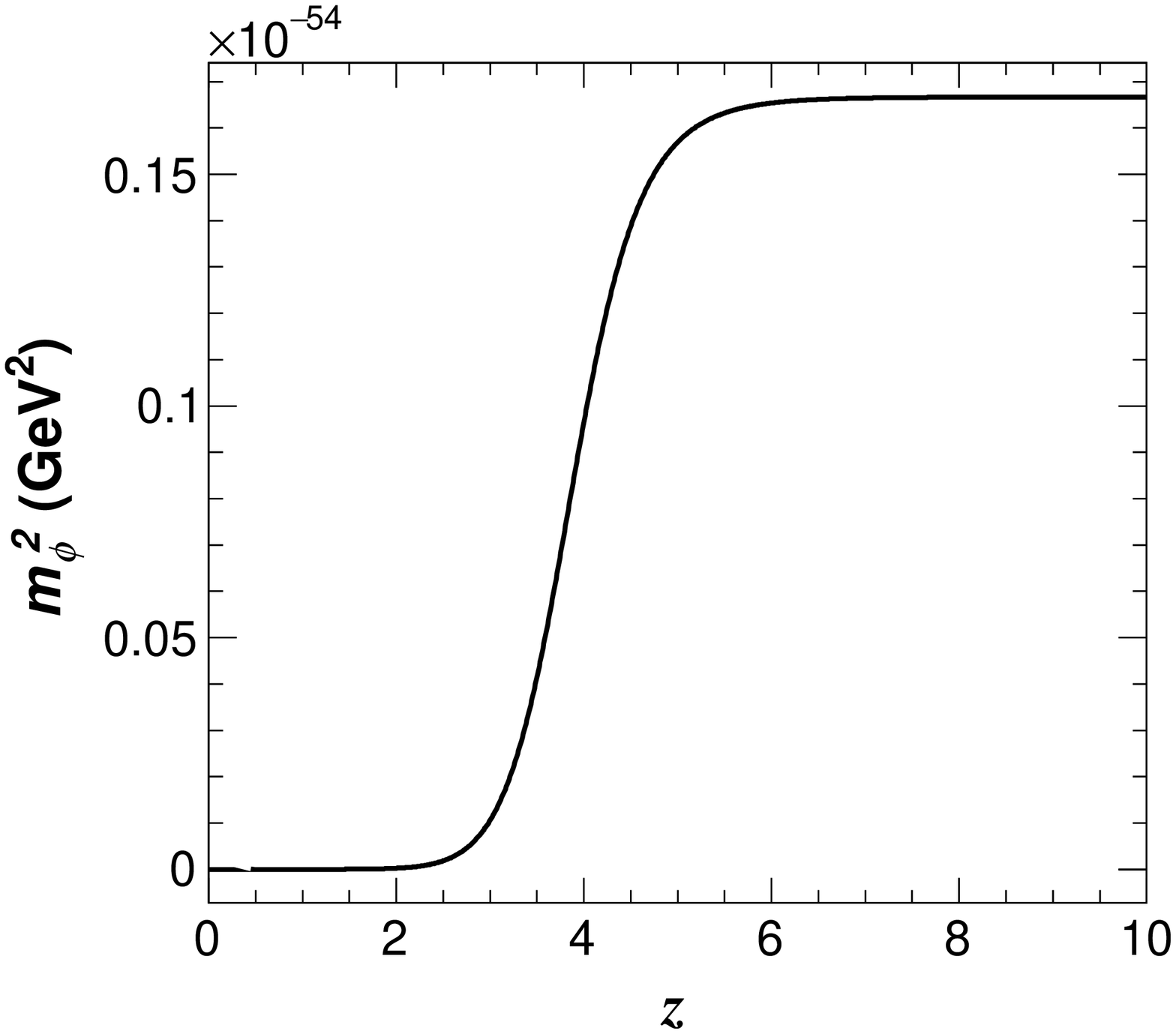} \hspace{0.2cm}
\includegraphics[scale = 0.28]{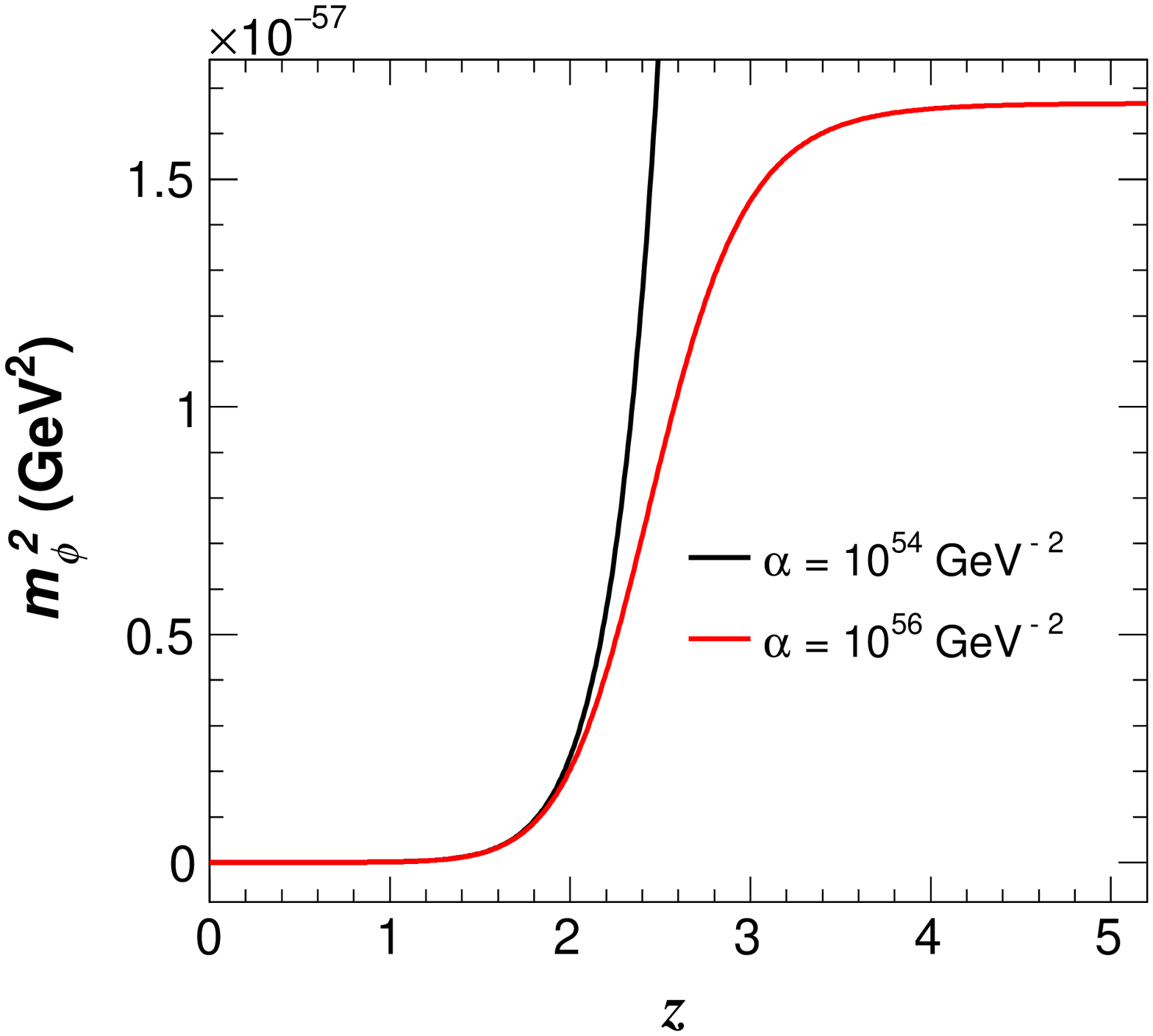}
}
\caption{Variation of scalaron mass in the Hu-Sawicki model as a function of 
redshift $z$ for the parameter $n = 2.5$. First two plots are for 
$\alpha = 10^{54}$ and the last plot is for two values of $\alpha$ as shown.
All these plots are drawn for the present average matter density of the 
Universe $\bar{\rho}_0 = 4\times10^{-25}$g/cm$^3$.}
\label{fig.7}
\end{figure}

%\begin{figure}
%\includegraphics[scale = 0.35]{mass_z_c2.eps}
%\caption{Variation of scalaron mass $m_{\phi}^2$ for the Hu-Sawicki model 
%as a function of redshift $z$ for $\alpha = 10^{25}$ GeV$^{-2}$ and 
%$\alpha = 10^{72}$ GeV$^{-2}$.}
%\label{fig.8}
%\end{figure}

The variation of scalaron mass as a function of redshift for different 
parameter values is shown in Fig.~\ref{fig.7}. 
The first two panels are for $\alpha = 10^{54}$GeV$^{-2}$ and the last panel is drawn
for two values of $\alpha$, $\alpha = 10^{54}$ and $10^{56}$ GeV$^{-2}$. The parameter
$n = 2.5$ and the present average matter density of the Universe 
$\bar{\rho}_0 = 4\times10^{-25}$g/cm$^3$ are used for all these plots. From 
this figure, it can be seen that the scalaron mass gradually increases with 
redshift, but this rate of increment is different at different stages of the 
Universe's evolution. That is, at high redshifts, the mass of the scalaron 
is also high. This means that with expansion of the Universe, the scalaron 
mass becomes light. Thus the scalaron was very heavy in the very early
Universe in comparison to its present range of values as mentioned already. 
This is in accordance with the chameleon mechanism. Again it can be seen 
that the scalaron mass increases with redshift upto a particular $z$ value
and then the curve becomes flat. This gives us an interesting result in 
analogy with the matter-dominated phase of the Universe. Moreover, for both 
the values of $\alpha$ the scalaron mass increases with redshift in a much 
similar fashion, however the lower value of $\alpha$ gives heavier scalaron
than that given by its higher value.

\section{Summary and Conclusions}
\label{sec.6}

In this paper, we have discussed about the scalaron, which is a new scalar 
field introduced from the modification of gravity via Weyl transformation of 
the spacetime metric $g_{\mu\nu}$ from the Jordan frame to the Einstein frame.
We have assumed that the fluctuation of the scalaron around the potential 
minimum can be regarded as dark matter.

In the first part, we discuss the $\mathit{f(R)}$ gravity which is one of the 
simplest modified gravity theories, using metric variational approach. We 
found that the Weyl transformation from the Jordan frame to the Einstein frame
adds a new degree of freedom. Then we obtained the potential of the scalaron 
which helped us to calculate the effective potential of the scalaron. We 
calculated the mass of the scalaron with a matter effect $T_\mu^\mu$. We found 
that the mass of the scalaron is dependent on the matter contribution.

In the second part, the chameleon mechanism is studied in the framework of the
Hu-Sawicki model of $\mathit{f(R)}$ gravity. The mass of the scalaron field 
for the Hu-Sawicki model shows its relation with the matter-energy density 
via the property known as chameleon mechanism, which is a screening mechanism 
for solar system tests of $\mathit{f(R)}$ gravity. For increasing values of the
model parameter $n$, the scalaron mass increases rapidly. Here, we found the 
mass of the scalaron to be very large. This large mass is a consequence of the
singularity problem that appears in Hu-Sawicki model of $\mathit{f(R)}$ 
gravity. So, to obtain a smaller mass of the scalaron field and to deviate 
from the singularity problem, we implemented the higher curvature correction 
to the original Hu-Sawicki model. As a result of the $R^2$ correction, we 
found the scalaron mass to be small in the large curvature region. Also, the 
singularity is pushed to infinity so that it is not easily attainable.

Furthermore, we discussed the properties of the scalaron field in the present 
Universe. The scalaron mass was calculated to be very light $\sim 10^{-23}$ 
eV, and satisfies the upper bound for the scalaron mass $m_\phi < \mathcal{O}(1)$ GeV. 
This mass of the scalaron field is quite closer to the mass of ultralight 
axion, which has an approximate mass scale of $\sim 10^{-22}$ eV. This 
ultralight axion is referred to as fuzzy cold DM. However, studies have shown 
that this low mass does not create any issue. Infact, it can be said that 
this scalaron makes small contributions to the amount of DM in the current 
Universe. Moreover, to explain observations of accelerated expansion much 
lighter scalar fields ($m \lesssim 10^{-33} \text{eV}$) are permitted by 
experimental evidences. Thus, we can come to the conclusion that the 
Hu-Sawicki model of $\mathit{f(R)}$ gravity can explain dark matter. 
In this work, we have implemented the pressureless dust model and hence, 
the scalaron (or axion) energy density does not scale as $\rho_\phi \sim 
a^{-3}$ and consequently it does not satisfy the perfect fluid continuity 
equation.

Finally, the evolution of the scalaron mass with the cosmological redshift is 
examined. It is seen that the mass of the scalaron increases with redshift 
upto a particular $z$ value and then becomes constant. It is known that the 
most distant objects manifest larger redshifts. Thus, our study shows that in 
the early Universe the scalaron was heavy and with expansion of the Universe, 
the scalaron becomes light. This is in accordance with the chameleon mechanism. 
Moreover, the rate of increase of scalaron mass with the redshift within the 
range of effective redshift values is different at different stages of 
evolution of the Universe.

Lastly, it should be mentioned that it would be interesting to extend further 
our work to study the interaction of the scalaron with SM particles. Another 
interesting aspect would be to apply this scalaron DM in other fields of 
particle physics to resolve different issues. Also, in near future we plan to 
work on other viable models of $\mathit{f(R)}$ gravity to study DM. Further 
possibility is considering a scalar field which can be a perfect fluid along 
with the chameleon used in our theory. Such $\mathit{f(R, \phi)}$ theories 
have been studied in \cite{oikonomou2021, Oikonomou2021}. Since dark matter 
can have multiple components, this would be an interesting future work.

\section*{Acknowledgments}
UDG is thankful to the Inter-University Centre for Astronomy and Astrophysics
(IUCAA), Pune for hospitality during his visits to the institute under the
Visiting Associateship program.

\end{document}